\documentstyle[prc,aps,epsfig,multicol]{revtex}
\draft
\normalbaselines
\def\bm {\bibitem}
\def\be {\begin{equation}}
\def\ee {\end{equation}}

\def\bea {\begin{eqnarray}}
\def\eea {\end{eqnarray}}

\begin{document}
\title{$\omega$ meson propagation 
in dense nuclear matter
and
collective excitations  
}
\author{ Abhee K. Dutt-Mazumder}
\address{TRIUMF Theory Group, 4004 Wesbrook Mall, Vancouver, BC, 
Canada V6T 2A3} 
\maketitle
\begin{abstract}
The bosonic excitations induced by the $\omega$ meson propagation 
in dense nuclear matter is studied within the framework of random phase
approximation. The collective modes are then 
analyzed by finding the zeros of the  relevant dielectric functions.
Subsequently we present closed form analytical expressions for the 
dispersion relations in different kinematical regime. Next, 
the analytical behaviour of the in-medium effective propagator for
the $\omega$ meson is  examined. This is exploited to calculate 
the full spectral function for the transverse (T) and longitudinal (L)
mode of the
$\omega$ meson. In addition, various sum rules are constructed for the
$\omega$ meson spectral density in nuclear medium. Results are then
discussed by calculating the residues at the poles and discontinuities 
across the cuts. 
\end{abstract}

\vspace{0.3 cm}
\section{Introduction}
One of the cardinal challenges facing the nuclear physics today is to
determine the properties hadrons at finite temperature and/or
chemical potential.  
Such studies are important both in the context of 
high energy heavy ion collisions (HIC) and  fixed target electron scattering 
experiments \cite{rappreview,cebaf}. Medium effects have also been found to
be important
to explain pion induced pion production data. It is seen that a  clear 
enhancement of $\pi^+\pi^-$ pair in the low invariant mass region
could be accounted for by invoking the in-medium $\sigma$ meson spectral function
\cite{hatsuda99,hirschegg00}. 
Similarly it is observed that 
the photoabsorption data in $\gamma + A$ system can 
be explained by invoking the medium modified $\rho$ meson spectral function
\cite{rappreview}. 

In heavy ion collision, among others, the light vector mesons have acquired 
particular interest. This is because of their dileptonic decay
channels. Dileptons, once produced leaves the medium without 
suffering further scattering and hence provide a penetrating probe to
study the properties of the dense nuclear medium as has been discussed at
length in the recent literatures{\cite{rappreview}. 
Experimentally the dilepton invariant mass spectra measured at CERN/SPS 
suggests that in-medium properties of the vector mesons, in particular that
of $\rho$ meson, are crucial to understand the low mass enhancement of the 
lepton pairs.

In the theoretical
arena such studies were stimulated by the pioneering work of Brown and 
Rho \cite{brown91}
where it was proposed that, in hot and/or dense matter, vector meson
masses would drop from their values in free space as a precursor phenomenon
of chiral symmetry restoration. Since then, several theoretical models have 
been invoked to study the in-medium properties of $\omega,\rho$ or $\phi$ 
meson which include QCD sum-rule, 
Nambu-Jona- Lasinio model, Chiral perturbation theory or simple hadron 
effective Lagrangian approach
\cite{jean94,shiomi94,hosaka90,saito98,abhee96}. 

In the present work we discuss the vector meson propagation in nuclear
matter which could be visualized as meson waves traveling through  
dispersive medium. This concept is usually invoked to study the
poinic collective effects in nuclear matter \cite{weise}.
A characteristic feature of meson propagation in matter is that
there exists two distinct scales depending upon its momenta  ; if the
momenta of the meson is small compared to the Fermi momenta of the nuclear
matter, the meson propagation becomes inseparable from the macroscopic 
collective oscillation of the system induced by the mesonic field, 
while on the other hand, for large momenta,
as we shall see, the meson propagation can still be regarded as usual particle
propagation with its dispersion characteristics very similar to that
of free space. In particular we study the propagation of the $\omega$ meson
in dense nuclear matter in the long-wave length limit i.e. in the regime
when the typical characteristic length over which the 
meson field varies ($1/q$) is large compared to the
the de Broglie wave-length of the nucleons  $1/p_F$. Such a study was
initially pursued by Chin \cite{chin77} who showed that, 
in nuclear matter, $\omega$ meson  picks up the collective oscillation 
of the system and becomes massive. However, in Ref.\cite{chin77}, 
the vacuum contribution (Dirac sea) was neglected on the ground that it 
only renders the couping constant momentum dependent and
in general, in the long-wave length limit ({\em i.e.} for low momentum 
transfer), the effect is only marginal. 
Later, in a seminal paper by Jean {\em et al.} \cite{jean94}
it was shown that, in presence of a mean scalar field 
(Walecka model), the vacuum part contributes substantially with opposite
sign. This, therefore, plays a crucial role in determining the over all
sign and magnitude of the net polarization insertion \cite{jean94}. 
In the current work we include both the effects and present new and more 
complete results for the $\omega$ meson dispersion relations in nuclear 
matter by considering various kinematical limits of the 
$\omega$ meson momentum compared to the density characterized by the Debye
mass or the plasma frequency of the system. Moreover, we also recover
the results presented in Ref. \cite{chin77} in the appropriate limits.

In addition to the calculation of the dispersion relations, we also
examine the analytical properties of the effective $\omega$ meson 
propagator for the longitudinal and transverse excitations
and construct various sum rules for the relevant spectral densities 
$(\rho_{T,L})$. Later, the full spectral functions $(\rho_{T,L})$ are expressed
in terms of the residues at the poles and cuts of the propagators. 
Such studies, as we shall see,  provide insight to understand and interpret 
the collective excitations of the nuclear matter via the sum rules and
the residues in different kinemetic domains of the momenta.

The paper is organized as follows. We first outline the formalism and 
describe the collective modes in terms of the zeros of the dielectric 
functions involving density dependent $\omega$ meson polarization
tensor. In section III, the concept of hard nucleon loop (HNL) approximation
is introduced which allows for the analytical evaluation of the
polarization insertions.
The dispersion relations for various kinematical domains are
then presented.
Subsequently, in section IV,
several sum rules for the $\omega$ meson spectral functions are constructed
by exploiting the properties of dispersion relation techniques of complex
variables.  Finally the conclusion and summary is presented in section V.

\section {Formalism}

It is well known that the collective modes corresponding to density 
fluctuations are determined from the poles of the meson propagators.  
In Ref.\cite{chin77} it has been demonstrated that in nuclear matter
such collective modes are dominated solely by the
vector interaction in the high-density limit. This, in turn, implies that
the in-medium properties of the vector mesons are directly linked with
the collective excitation set by their propagation.

The dielectric functions relevant for the $\omega$ meson propagation in
dense nuclear matter can be calculated within the frame work of random
phase approximation (RPA) which essentially implies repeated insertion of
the $\omega$ meson self-energy involving nucleon-nucleon loop as described 
in ref.\cite{chin77,serot86}. 

Considering the interaction
of the $\omega$ meson with the nucleon field,  
${\cal L}_I=g_V{\bar\psi}\gamma_\mu\psi\omega^\mu$, 
the second order polarization tensor $\Pi_{\mu\nu}$,
can be written as   
\begin{equation}
\Pi_{\mu\nu}^{\alpha\beta}=\frac{-i}{(2\pi)^4}
\int {d^4}p~{\rm Tr}[i\gamma^\alpha_\mu iG(p+q)
{i\gamma^\beta_\nu} iG(p)]
\label{eq:pimunu}
\end{equation}
where $(\alpha,\beta)$ are the isospin indices and $G(p)$ is the
in-medium nucleon propagator\cite{serot86}

\bea
G(p)=G_F(p) + G_D(p)
\label{eq:propfull}
\eea

where

\begin{equation}
G_F(p) = 
(p_\mu \gamma^\mu +M^*)[
\frac{1}{p^2-M^{* 2}+i\epsilon} ]
\end{equation}
and
\begin{equation}
G_D(p) = 
(p_\mu \gamma^\mu +M^*)[
 \frac{i\pi}{E^*(p)}\delta (p_0-E^*(p))\theta(p_F-|\vec p |)].
\end{equation}
The first term in G(p), namely, $G_F(p)$, is the same as the free propagator
of a spin 1/2 fermion, while the second part, $G_D(p)$, involving
$\theta(p_F-|\vec p |)$, arises from Pauli blocking, describes the modification
of the same in the nuclear matter at zero temperature. $E^*(p)=\sqrt{p^2
+ M^{* 2}}$ and  $M^*$ denotes the
in-medium mass of the nucleon, which in the present context might be 
determined from the mean scalar density \cite{serot86}.

In a similar vein the polarization insertions can also be written as sum of
two parts :
\bea
\Pi_{\mu\nu}=\Pi_{\mu\nu}^F + \Pi_{\mu\nu}^D
\label{eq:polten0}
\eea.

From the above set of equations we can write the real part of the 
density dependent 
$\omega$-meson self energy as
\cite{chin77,abhee96} 
\bea
\Pi_{\mu\nu}^D&=&\frac{g_v^2}{\pi ^3}\int_0^{p_F}\frac{{d^3p}}{E(p)}
\frac{{\cal P}_{\mu\nu} q^2 -Q_{\mu\nu}(p\cdot q)^2}{q^4-4(p\cdot q)^2} 
\label{eq:polten}
\eea

\noindent where ${\cal P}_{\mu\nu}=(p_\mu-\frac{p.q}{q^2}q_\mu)
(p_\nu-\frac{p.q}{q^2}q_\nu)$ and $Q_{\mu\nu}=(-g_{\mu\nu} +
\frac{q_{\mu}q_{\nu}}{q^2})$. It is evident that the form of the 
polarization tensor conforms to the requirement of current conservation,
i.e. $ q^\mu\Pi^D_{\mu\nu}=\Pi^D_{\mu\nu}q^\nu=0$
In order to evaluate $\Pi^D_{\mu\nu}$ conveniently, we choose $\vec q$ to be
along the $z$ axis i.e. $ q=(q_0,0,0,\mid\vec q\mid) $, and 
$p.q=E(p)q_0 -\mid p\mid \mid q \mid \chi$, 
where $\chi$ is the cosine of the
angle between $\vec k$ and $\vec q$. On account of integration over the
azimuthal angle, only six components survive. Then again
for isotropic nuclear matter we have
$\Pi_{11}^D=\Pi^D_{22}$ and $\Pi^D_{03}=\Pi^D_{30}$. Furthermore, the
current conservation condition puts additional constraints leaving
only two non-vanishing
independent component of $\Pi^D_{\mu\nu}$, linear combinations of
which gives us the longitudinal and transverse components of
$\Pi^D_{\mu\nu}$, namely, $\Pi^D_L(q)=-\Pi_{00}^D+\Pi_{33}^D(q)$
 and $\Pi_T^D(q)=\Pi^D_{11}=\Pi^D_ {22}$ 

The free part is obtained by putting $G(p)=G_F(p)$ in Eq.~\ref{eq:pimunu}
which is divergent and can be regularized using appropriate renormalization 
scheme \cite {chin77,hatsuda96}.  The regularization procedure
we adopt here is the following :
\bea
\partial^n\Pi^F(q^2)/\partial(q^2) ^n\vert _{M^\ast_n\rightarrow M,
q^2=m_\omega^2}=0, 
\eea
where, $ (n=0,1,2...,\infty )$. This yields
\bea
\Pi^F_{\mu\nu}=
Q_{\mu\nu}\frac{g_v^2}{\pi^2} q^2 \int_0^1 dx x(1-x) 
log[\frac{M^{* 2} - q^2 x (1-x)} {M^{2} - m_\omega^2 x (1-x)}].
\label{eq:pif0}
\eea
It is evident that free part does not distinguish between the longitudinal
and transverse mode and we can take $\Pi^F_L=\Pi^F_T=\Pi_{22}=\Pi_{11}$.
In order to study collective response of the nuclear system and to obtain 
the dispersion relation of the induced oscillation, one needs to find
zeros of the dielectric function defined through 
\bea
\epsilon_{T,L}(q_0,|q|)=
1-\frac{1}{q_0^2-|q|^2- m_\omega^2}\Pi_{T,L}(q_0,|q|)=0,
\label{eq:epsilon}
\eea
where $\Pi_{T,L}(q_0,|q|)=\Pi^F_{T,L}(q^2)+ \Pi^D_{T,L}(q_0,|q|)$. 
It is to be noted that the free part $\Pi^F_{L,T}$ depends only
on the Lorentz scalar $q^2$ while the density dependent part
$\Pi_{L,T}^D$ involves both $q_0$ and $|q|$ individually.
The equations $\epsilon_{T,L}(q_0,|q|)=0$ in general have to be solved 
numerically. However, in the next section,
we discuss appropriate approximation scheme which admits close form 
solutions.
 
\section{Collective Modes and Dispersion relations}

In this section we discuss first the approximation scheme to derive
the polarization tensors analytically.  Once the polarization insertions are 
determined one can solve Eq.~\ref{eq:epsilon} to find the dispersion relations 
for the $\omega$ meson in nuclear matter taking both the effect of free and
dense part of the polarization function in the various kinematic limits as 
mentioned in the introduction.  In addition, this also provides us a way to 
interpret the results in terms of the classical plasma oscillations. 

First we recall that for collective excitations, the wavelength of the 
oscillations must be greater than the interparticle spacing. This means
the meson momenta must be small compared to the nucleon momenta. Therefore,
quantitatively, it is legitimate to assume  that all the loop momenta 
(nucleon) are hard and the external momenta (meson) are soft {\em i.e} 
$p\sim p_F$, $|q|<<p_F$.  We term this as hard nucleon loop (HNL) approximation.
This allows us to drop $q^4$  compared to $4(p.q)^2$ in Eq. ~\ref{eq:polten}. 
Such an approximation, is, in effect, similar to what has been 
adopted in Ref. \cite{chin77}.  Under this assumption the 
integrations of Eq.~\ref{eq:polten} 
relevant for the longitudinal  and transverse 
modes are performed easily to give following results :

\bea
\Pi^D_L(\omega,k)=
-3\Omega^2(1-\frac{\omega^2}{k^2v_F^2})
[-1+\frac{\omega Log[\frac{\omega+k v_F}{\omega-k v_F} ]}{2kv_F}]
\label{eq:pil}
\eea
and 
\bea
\Pi^D_T(\omega,k)=\frac{3}{2}\Omega^2[\frac{\omega^2}{k^2v_F^2}+
(1-\frac{\omega^2}{k^2v_F^2})
\frac{\omega Log[\frac{\omega+k v_F}{\omega-k v_F} ]}{2kv_F}]
\label{eq:pit}
\eea
where, $q_0=\omega$, $|q|=k$, $v_F=p_F/\epsilon_F$ is the Fermi velocity with  
$\epsilon_F=\sqrt{p_F^2+M^2}$ and
$\Omega^2$ is identified as the plasma frequency given by 
$\Omega^2=\frac{g_V^2}{\pi^2}\frac{1}{3}\frac{p_F^3}{\epsilon_F}$. 
This is related to the Debye screening mass as $\Omega^2=\frac{2}{3} m_D^2$
\cite{bellac,kapusta}.

Before proceeding further, few comments are in order. 
It is evident that for $k\neq 0$ in matter, the longitudinal and transverse
self-energies are non-degenerate. This gives rise to splitting between these
two collective modes in nuclear matter
which could be attributed to the presence of ${\cal P}_{\mu\nu}$ in 
Eq.~ \ref{eq:polten}.  However, they become identical in the static
limit {\em{i.e.}} 
$\Pi_L(\omega,k\rightarrow 0)=\Pi_T(\omega,k\rightarrow 0)=\Omega^2$. 

Next we consider the free part (also known as Dirac sea contribution) given
by Eq.~\ref{eq:pif0}.  Following Ref.\cite{jean94}, this can also be simplified 
further by making a Taylor series expansion of 
Eq.~\ref{eq:pif0} around $M^*$.  To the leading order the contribution is
found to be \cite{jean94,thomasprivate}
\bea
\Pi^F_{L,T}(\omega,k)=
\frac{g_v^2}{3\pi^2}\frac{M^*-M}{M} m_\omega^2 
+ {\cal O}(m_\omega^2/4M^2)
\label{pif}
\eea
As in nuclear matter effective nucleon mass is smaller than
its free space value ( $M^*<M$), it is clear from the last
expression, that the free part reduces the effective $\omega$ meson mass
in the medium.

Eq.~\ref{eq:epsilon} in conjunction with Eq.~\ref{eq:pit}, ~\ref{eq:pil} 
and ~\ref{pif} define the
the dispersion characteristics for the collective excitations: 

\bea
\omega^2-k^2-m_\omega^2 
-\frac{g_v^2}{3\pi^2}\frac{M^*-M}{M} m_\omega^2 
-\frac{3}{2}\Omega^2[\frac{\omega^2}{k^2v_F^2}+
(1-\frac{\omega^2}{k^2v_F^2})
\frac{\omega Log[\frac{\omega+k v_F}{\omega-k v_F} ]}{2kv_F}]
=0
\label{eq:epsT}
\eea
for T mode and 
\bea
\omega^2-k^2-m_\omega^2  
-\frac{g_v^2}{3\pi^2}\frac{M^*-M}{M} m_\omega^2 
+3\Omega^2(1-\frac{\omega^2}{k^2v_F^2})
[-1+\frac{\omega Log[\frac{\omega+k v_F}{\omega-k v_F} ]}{2kv_F}]
=0,
\label{eq:epsL}
\eea
for L mode.

Generally these transcendental equations, {\em viz.} 
Eq.~\ref{eq:epsT} and \ref{eq:epsL}, have to be solved numerically,
however, analytical solutions could be obtained in two limiting cases 
{\em viz.}, when $k<\Omega$ and $k>\Omega$ by expanding above equations
in powers of $k$ and  solving the equations by successive iterations. To
the first order, when $k<<\Omega$ one gets 
${\omega_{L(T)}^2=m_\omega^{*2}
+ \Omega^2}$. 
We note that the modes are degenerate in this limit and
oscillations are independent of the wave vector $k$. In 
the case of electron-ion plasma, this mode was identified as the 
`plasma waves' or `Langmuir' oscillation and $\Omega^2$ there is
given by $e^2\rho/m^2$ where $\rho$ is the density of electron gas 
and $m$ is the electron mass \cite{lifshitz}. 
This justifies identifying $\Omega$ as the plasma 
frequency in the present pretext. It should be mentioned that
here we have absorbed the contribution of the free part as 
\bea
m_\omega^{* 2}= m_\omega^2 ( 1 +  \frac{g_v^2}{3\pi^2}\frac{M^*-M}{M} )
\eea.

Once the leading order solution 
is known, we can solve Eqs.~\ref{eq:epsT} and \ref{eq:epsL}
iteratively to obtain

\bea
\omega_T^2&=&m_\omega^{* 2} + \Omega^ 2+ k^2 + \frac{1}{5}k^2 v_F^2 
\frac{\Omega^2}{m_\omega^{* 2}+ \Omega^ 2} + .... 
\label{eq:disprelT}
\\
\omega_L^2&=&m_\omega^{* 2}+ \Omega^2 + \frac{3}{5}k^2 v_F^2 
\frac{\Omega^2}{m_\omega^{* 2}+\Omega^ 2} + ....
\label{eq:disprelL}
\eea

Evidently, the transverse mode lies above the longitudinal one and it
requires more energy for excitation. However, the splitting in the
low $k$ region is very marginal. In the limit $m_\omega\rightarrow 0$ we recover the results presented 
in \cite{chin77}, {\em i.e.} 
\bea
\omega_T^2&=&\Omega^2 + k^2 + \frac{1}{5} v_F^2 k^2 + ...\\
\omega_L^2&=& \Omega^2 + \frac{3}{5} v_F^2 k^2 + .....
\eea
Furthermore, this taken with $v_F\rightarrow 1$ results in 
dispersion relations arising out of the photon propagation at finite 
temperature or density (where electron is considered to be ultrarelativistic)
with the appropriate definition of the plasma frequency 
\cite{kapusta,blaizot02}. 
It is worthy to note that the contribution of the density dependent part
is opposite to that of the Dirac part of the polarization tensor. This is
consistent with Ref.\cite{jean94}. 

Next we consider the case when $k>\Omega$ but still $\omega > k$ and in fact
$\omega \sim k v_F$. 
The successive approximation for this asymptotic value of $k$ yields solutions 
of the form

\bea
\omega_T^2=k^2+m_\omega^{ * 2} + \frac{3}{2}\Omega^ 2 + ...
\eea

The corresponding longitudinal frequency takes little bit complicated 
form;
\bea
\omega_L^2=k^2v_F^2 ( 1 + 
4~ exp( -\frac{2}{3}\frac{k^2v_F^2}{\Omega^2} 
+ \frac{m_\omega^{ * 2}}{3\Omega^2}\frac{v_F^2}{(v_F^2-1)}-2
 ))
\eea

Evidently, the longitudinal mode approaches the line $\omega=k v_F$ and
exponentially suppressed for very large values of $k$. We have solved 
Eq. \ref{eq:epsT} and \ref{eq:epsL} numerically and checked that for
meson momenta $k\sim 1 GeV$ dispersion results are in good agreement with
Eqs. \ref{eq:disprelT} and  \ref{eq:disprelL}.

\section{Spectral functions and sum-rules}

Equipped with Eq.~\ref{eq:pil} and ~\ref{eq:pit}, we can 
proceed to calculate the spectral functions of the $\omega$ meson.
As demonstrated already, the $\omega$ meson in matter has two modes. 
Confining our attention 
on the transverse sector  first, we write the propagator as

\bea
\Delta_T(\omega,k)
&=&\frac{-1}{\omega^2-k^2-m_\omega^{ * 2} -\frac{3}{2}{\Omega^2}[
{\frac{\omega^2}{(kv_F)^2} + (1-\frac{\omega^2}{(kv_F)^2})
\frac{\omega}{2 kv_F}Log[\frac{\omega + k v_F}{\omega-k v_F}}
]]
}
\label{eq:tprop}
\eea
Next chain of steps would be to construct the sum rules by 
exploiting the analytic properties this effective propagator. First we
note that the function $\Delta_T$ is analytic in the complex $\omega$-plane
having a cut from $-kv_F$ to $+ kv_F$; in addition to the poles 
$\omega=\pm\omega_T$. Having observed this one can  make use of
Cauchy's theorem for $\Delta=\Delta_T$
\bea
\Delta(\omega,k)&=&\oint_\Gamma \frac{d\omega^\prime}{2\pi i} 
\frac{\Delta(\omega^\prime,k)}{\omega^\prime-\omega}\nonumber\\
&=&\int_{-\infty}^\infty \frac{d\omega^\prime}{2\pi i} 
\frac{\Delta(\omega^\prime+i\eta,k) - \Delta(\omega^\prime-i\eta,k)}
{\omega^\prime-\omega} +
\oint_{\Gamma^\prime} \frac{d\omega^\prime}{2\pi i}
\frac{\Delta(\omega^\prime,k)}{\omega^\prime}
\eea
In the above equation  $\Gamma^\prime$ is a circle with its radius pushed
to infinity. This equation can be rewritten in terms of the spectral density
$\rho(\omega,k)$ containing both the discontinuities across the cuts and
the residues at the poles:

\be
\rho(\omega,k)=2 Im \Delta(\omega+i\eta,k)
\ee
as

\bea
\Delta(\omega,k)
&=&\int_{-\infty}^\infty \frac{d\omega^\prime}{2\pi } 
\frac{\rho(\omega^\prime,k)}
{\omega^\prime-\omega} +
\oint_{\Gamma^\prime} \frac{d\omega^\prime}{2\pi i}
\frac{\Delta(\omega^\prime,k)}{\omega^\prime}
\label{eq:spec_reprT}
\eea

It is evident from Eq.~\ref{eq:tprop} that $\Delta_T(\omega^\prime,k)\rightarrow
\frac{1}{\omega^{\prime 2}}$ as $\omega^\prime\rightarrow \infty$, therefore,
the integration over the contour $\Gamma^\prime$ fails to contribute in this
region.  First sum rule is then immediately obtained by setting $\omega=0$
in Eq.~\ref{eq:spec_reprT}

\bea
\int_{-\infty}^\infty\frac{d\omega}{2\pi}\frac{\rho_T(\omega,k)}{\omega}
&=&\Delta_T(0,k)=\frac{1}{k^2+m_\omega^{ * 2}}
\label{srt_0}
\eea

Other sum rules are obtained by examining the asymptotic behaviour 
of $\Delta_T(\omega,k)$ when 
$\omega \rightarrow  \infty$; by taking the parity property of the spectral
density {\em i.e.} $\rho_T(\omega,k)=-\rho_T(-\omega,k)$ into account, one can
write from Eq.~\ref{eq:spec_reprT} for $\omega\rightarrow \infty$

\bea
\Delta_T(\omega,k)=-\frac{1}{\omega}\sum_{n=0}^\infty
\int_{-\infty}^\infty
\frac{d\omega^\prime}{2\pi}(\frac{\omega^\prime}{\omega})^{2n+1} 
\rho_T(\omega^\prime,k)
\label{eq:tprop_asymp}
\eea

On the other hand, one can expand Eq.~\ref{eq:tprop} in powers of
$\omega^{-1}$; comparing this with  Eq.~ \ref{eq:tprop_asymp} one 
writes for $n=0$
\bea
\int_{-\infty}^\infty\frac{\omega d\omega}{2\pi}\rho_T(\omega,k)=1
\label{eq:srt_1}
\eea
and similarly for $n=1$ one observes
\bea
\int_{-\infty}^\infty\frac{\omega^3 d\omega}{2\pi}\rho_T(\omega,k)
=k^2+m_\omega^{ * 2}+\Omega^2
\label{eq:srt_2}
\eea
Thus we obtain other two sum rules. Of particular importance is 
Eq.~\ref{eq:srt_1}. This does not depend on $k$ which could be shown
to be a consequence of the canonical commutation relation. 
Therefore,
it is natural to expect that such a relation will also be satisfied by the
transverse $\omega$ meson in nuclear matter. 

%%% Longitudinal Mode Sum Rules
The longitudinal  propagator is denoted as  
\bea
\Delta_L(\omega,k)&=&
\frac{\omega^2-k^2}{k^2}
\frac{-1}{\omega^2-k^2 - m_\omega^{ * 2} +
3\Omega^2(1-\frac{\omega^2}{k^2v_F^2} )[-1+
{
\frac{\omega}{2 kv_F}Log[\frac{\omega + k v_F}{\omega-k v_F}}
]]
}
\eea
Contrary to the case of transverse propagator, we now have non-vanishing
contribution from the contour denoted by $\Gamma^\prime$ in 
Eq. ~\ref{eq:spec_reprT} as in the limit
${\omega\rightarrow\infty}$ the logarithmic term $\rightarrow 1$.
Hence in this limit the longitudinal propagator goes like 
$\Delta_L(\omega,k) \rightarrow  -1/k^2 $.
Therefore, a subtraction, unlike the transverse propagator, is necessary. 
The subtracted dispersion relation gives the following sum rule essentially
in the same manner as in the transverse case. 
\bea
\int_{-\infty}^\infty\frac{d\omega}{2\pi} \frac{\rho_L(\omega,k)}{\omega}
&=& \frac{1}{k^2}+\Delta_L(0,k) = 
\frac{3\Omega^2+m_\omega^{ * 2}}{
k^2 ( k^2+m_\omega^{ * 2} + 3 \Omega^2 )
}\\
\eea
Similarly to the asymptotic behaviour of the longitudinal propagator
one can write,
\bea
\Delta_L(\omega,k)=-\frac{1}{k^2} -\frac{1}{\omega}\sum_{n=0}^\infty
\int_{-\infty}^\infty
\frac{d\omega^\prime}{2\pi}(\frac{\omega^\prime}{\omega})^{2n+1} 
\rho_T(\omega^\prime,k)
\label{eq:lprop_asymp}
\eea
This is identical with Eq.~\ref{eq:tprop_asymp} except here we have an
additional factor of $-1/k^2$ for reason mentioned in the previous 
paragraph. Likewise, we now can derive the sumrules by first setting 
$n=0$ to get
\bea
\int_{-\infty}^\infty\frac{\omega d\omega}{2\pi}\rho_L(\omega,k)=
\frac{m_\omega^{ * 2}+\Omega^2}{k^2}
\label{srl_1}
\eea

and $n=1$ to obtain
\bea
\int_{-\infty}^\infty\frac{\omega^3 d\omega}{2\pi}\rho_L(\omega,k) =
\frac{
(\Omega^2+m_\omega^{ * 2})
(k^2+m_\omega^{ * 2}+\Omega^2)
-\frac{2}{5}k^2\Omega^2 v_F^2}{k^2}
\label{srl_2}
\eea

We can write down the full spectral function of the transverse and
longitudinal modes for the $\omega$ meson in nuclear matter in terms
of the poles (and residues at the poles) together with the cuts of the
propagator. 

\bea
\frac{1}{2\pi}\rho_{T,L}(\omega,k)=Z_{T,L}(k)[
\delta(\omega-\omega_{T,L})
-\delta(\omega+\omega_{T,L})] + \beta_{T,L}(\omega,k)\theta(k^2-\omega^2)
\eea

where, $\beta_{T,L}$ is given by
\bea
\beta_{T,L}(\omega,k)=  \frac{Im\Pi_{T,L}}{\pi} 
\vert\Delta_{T,L}(\omega,k)\vert^2
\eea

The imaginary part of the self-energies 
$\Pi_{T,L}$ correspond to Landau damping and
relevant only for the space-like momenta of the meson. Physically this
refers to the energy loss due to scattering of the $\omega$ meson with 
hard nucleons in the medium. They are given by the imaginary part of 
Eq.\ref{eq:pil} and Eq.\ref{eq:pit}. 

The residues at the poles which are used
to determine the full spectral function of the
$\omega$ meson in nuclear matter can be determined from the one-loop
effective propagators. For instance the residue 
at the pole for the transverse excitation is determined by
\bea
Z_{T}
&=&-([\partial\Delta_T^{-1}/\partial\omega]_{
\omega=\omega_{T(k)})}])^{-1}\\
&=&\frac{\omega_T ( k^2 v_F^2 - \omega_T^2)}{\omega_T^2(\omega^2-3
(\Omega_T^2+k^2+m_\omega^{* 2}))+ k^2(k^2+\omega_T^2
+m_\omega^{* 2})v_F^2}
\label{eq:resT}
\eea

In a similar way we also calculate the reside at the pole for the longitudinal
oscillation which is found to be 
\bea
Z_L
&=&-\frac{(k^2-\omega_L^2)(-\omega_L^3+\omega_L k^2 v_F^2)}{k^2 
[\omega_L^2(-3 k^2 - 3 m_\omega^{* 2} + \omega_L^2 - 3 \Omega^2   )
+ k^2( k^2 + m_\omega^{* 2} + \omega_L^2 + 3\Omega^2) v_F^2]}
\label{eq:resL}
\eea

We now present the limiting values of these residues at the poles given in
section III. First considering $k<\Omega$, we have for the transverse 
$\omega$
\bea
Z_{T}(k)\simeq
\frac{1}{2 \sqrt{\Omega^2+m_\omega^{* 2}}}
-\frac{k^2}{20}
(\frac{5 m_\omega^{* 2} + ( 5 + 3 v_F^2)\Omega^2}{(\Omega^2+m_\omega^{* 2})
^{5/2}})
\label{zt0}
\eea
while for large k, {\em i.e.} for $k>\Omega$ we have

\bea
Z_{T}(k)\simeq
\frac{1}{2 \sqrt{k^2 + m_\omega^{* 2}}}.
\label{zt1}
\eea

Similar results are obtained for the longitudinal mode also. They are given
by, for $k<\Omega$

\bea
Z_{L}(k)\simeq  
\frac{\sqrt{\Omega^2+m_\omega^{* 2}}}{2 k^2 }
- \frac{ 25(m_\omega^{* 2}+ \Omega^2) - 22 v_F^2 \Omega^2 }
{20 (\Omega^2+m_\omega^{* 2})^{3/2} }
\label{zl0}
\eea

and for $k>\Omega$
\bea
Z_{L}(k)\simeq  \frac{4 k v_F}{3\Omega^2} 
exp( -\frac{2}{3}\frac{k^2v_F^2}{\Omega^2} 
+ \frac{m_\omega^{ * 2}}{3\Omega^2}\frac{v_F^2}{(v_F^2-1)}-2
 )
\label{zl1}
\eea

It is now evident from Eq. 28 and Eq.~\ref{zt1}
 that the transverse spectral function for large meson momentum
reduces to that of the $\omega$ meson propagation in free space, except its
mass is modified because of the presence of the mean scalar field. Longitudinal
mode in this limit decouples from the system.  For small momenta, the
sum rules are almost saturated by the pole contributions. 
This could be checked numerically. 

\section {Summary and Conclusion} 

To conclude and summarize, in the present work, we investigate the 
collective modes induced by the $\omega$ meson propagation in dense nuclear
matter. We then determine the dispersion relations for the $\omega$ meson
by considering both the effect of the Fermi sea and Dirac vacuum
within the scheme of HNL approximation. It is observed that in the 
appropriate limit, i.e.  for meson momenta small compared to the plasma 
frequency of the system, the results are consistent with the previous 
calculations \cite{jean94,chin77}. The
$\omega$ meson mass, in presence of a mean field, decreases even at normal
nuclear matter while for free Fermi gas the effect of nucleon loop is to
increase the $\omega$ meson mass proportional to the characteristic plasma
frequency of the system. 

We also present
analytical results for the dispersion relation in the limit $k>\Omega$
which shows that for the large value of the momenta the medium effects die off
and the transverse mode propagates like a free meson.  The longitudinal mode
in this limit decouples from the system. We also construct various
sum rules satisfied by the $\omega$ meson spectral density in nuclear matter
by investigating the analytic properties of the relevant
effective propagator at finite density. 
It is observed that that sum rules 
are mostly saturated by the pole contributions for small momenta. 
The residues are also presented in
the different limit of the meson momenta to understand the collective modes
and their behavior. 
Results pertinent to the classical electron plasma oscillations are also
recovered. 

Similar studies can also be pursued for other light vector mesons. For
instance, it is known that the $\rho$ nucleon interaction is very similar to 
the $\omega$ nucleon interaction, where, in addition to the vector interaction, 
one has to include tensor interaction as well. We, therefore, believe that the
$\rho$ meson propagation in cold nuclear matter can also be studied within
the framework HNL approximation. This would allow for an analytical derivation 
for the spectral density of the $\rho$ meson. Hence ,
sum rules similar to what we present here, relevant
for the $\rho$ meson spectral function at finite 
density can also be constructed in a parallel approach. Such studies are in 
progress and shall be reported elsewhere.

\section{Acknowledgments}
The author would like to thank A. G. Williams and J. Piekarewicz
for useful private communications.


\begin{references}

\bm{rappreview}
R. Rapp and J. Wambach, Adv. Nucl. Phys. {\bf 25},  1  (2000).

\bm {cebaf} B.M. Preedom and G.S. Blanpied, CEBAF report PR-89-001 (1989).

\bm {hatsuda99} T. Hatsuda, T. Kunihiro and H. Shimizu,
Phys. Rev. Lett {\bf 82}, 2840 (1999).

\bm{hirschegg00}  
Proceedings of the International Workshop XXVIII on
Gross Properties of Nuclei and Nuclear Excitations,
Ed. M. Buballa, W. Norenberg, B-J Schaefer and J. Wambach, 2000.  

\bm{brown91}
G.E. Brown and M. Rho, Phys. ReV. Lett, {\bf 66}, 2720 (1991)

\bm {jean94} 
H.-C. Jean, J. Piekarewicz, and A. G. Williams, 
Phys. Rev C, {\bf 49}, 1981 

\bm {shiomi94} H. Shiomi and T. Hatsuda, 
Phys. Lett. {\bf{B 334}}, 281 (1994).

\bm{hosaka90} A. Hosaka, Phys. Lett {\bf B244}, 363, (1990).

\bm{saito98} K. Saito, K. Tsushima, A. W. Thomas and A. G. Williams
Physics Letters {\bf B 433}, 243 (1998).

\bm {abhee96} A. K. Dutt-Mazumder, B. Dutta-Roy, A. Kundu and T. De,
Phys.Rev. C53, 790 (1996)

\bm {weise}T. Ericson and W. Weise, {\it Pions and Nuclei} (Oxford Sc. Publ.,
Oxford, 1998).

\bm {chin77} S. A. Chin, Ann. Phys. 108, 301 (1977)

\bm{serot86} Brian D. Serot and John D. Walecka, Adv.
Nucl. Phys. {\bf 16}, 1 (1986).

\bm {hatsuda96}
T. Hatsuda, S. H. Lee and H. Shiomi,
Prog.Theor.Phys. {\bf 95}, 1009 (1996) 

\bm{bellac} M. Le Bellac, Thermal Field Theory, Cambridge monographs
on mathematical physics, Cambridge University Press, 1996.

\bm{kapusta} J. I. Kapusta, Finite Temperature Field Theory, 
Cambridge University Press, Cambridge, 1989.

\bm {thomasprivate} A. G. Williams, Private Communication.

\bm{lifshitz} E. M. Lifshitz and L. P. Pitaevskii, Physical Kinetics, Vol. 10,
Pergamon Press, 1981.


\bm{blaizot02} J. P. Blaizot, E. Iancu, Phys.Rept.359, 355 (2002) 

\end{references}
\end{document}